\documentstyle[aps,multicol,epsf]{revtex}
\input epsf
\draft
\title{High resolution amplitude and phase gratings in atom optics}
\author{P. R. Berman, B. Dubetsky, and J. L. Cohen 
\\ {\em Physics Department, University of Michigan, Ann Arbor, MI 48109-1120} }
\date{\today }
\begin{document}
\maketitle
\begin{abstract}
An atom-field geometry is chosen in which an atomic beam traverses a field
interaction zone consisting of three fields, one having frequency $\Omega
=c/\lambda $ propagating in the ${\bf \hat{z}}$ direction and the other two
having frequencies $\Omega +\delta _{1}$ and $\Omega +\delta _{2}$
propagating in the -${\bf \hat{z}}$ direction. For $n_{1}\delta
_{1}+n_{2}\delta _{2}=0$ and $\left| \delta _{1}\right| T,\left| \delta
_{2}\right| T\gg 1$, where $n_{1}$ and $n_{2}$ are positive integers and $T$
is the pulse duration in the atomic rest frame, the atom-field interaction
results in the creation of atom amplitude and phase gratings having period $%
\lambda /[2(n_{1}+n_{2})]$. In this manner, one can use optical fields
having wavelength $\lambda $ to produce atom gratings having periodicity
much less than $\lambda $.
\end{abstract}
\begin{multicols}{2}
\narrowtext
\section{Introduction}

Over the past several years, atom interferometry has emerged as an important
new research area in atomic, molecular and optical physics \cite{atom
interferometry}. The technology has improved to the point where it is now
possible to control the center-of-mass motion of atoms using either
microfabricated gratings \cite{micro gratings} or optical fields \cite
{optical gratings}. Accompanying the developments in atom interferometry and
atom optics have been attempts to produce nanostructures using atom optics
elements. Probably the most successful method to date employs standing wave
optical fields to focus atoms to a series of lines or dots having size on
the order of tens of nanometers \cite{optical focusing}. The period of the
structures produced in these focusing schemes is $\lambda /2$, where $%
\lambda $ is the wavelength associated with the standing wave fields.
Spatial features having dimensions as small as $\lambda /8$ have been
achieved by exploiting the ground state optical potentials for a
magnetically degenerate ground state \cite{Gupta}, but the period of the
entire pattern remains equal to $\lambda /2.$

We have described previously a method for producing spatially modulated
atomic densities \cite{dubet1}. When an atomic beam passes through one or
more standing wave optical fields having wavelength $\lambda $, it is
possible to use coherent transient techniques to create atomic ``gratings''
having spatial period equal to $\lambda /(2n)$, where $n$ is a positive
integer. The atomic gratings arise as a result of the nonlinear interaction
between the atoms and the fields. For example, when an atomic beam passes
through a resonant, standing wave optical field, the field can create all
even spatial harmonics in the excited and ground state populations. As a
result of spontaneous emission, the excited state gratings decay back to the
ground state; however, for properly chosen level schemes the ground state
gratings are not completely ''filled in.'' As a consequence, one has
long-lived ground state gratings with which to operate. The fate of these
ground state gratings depends on the collimation of the atomic beam.

If the angular divergence of the atomic beam is $\theta _{b}$, the gratings
persist for a distance of order $L_{b}=\lambda /(2n\theta _{b})$ following
the atom-field interaction. For distances larger than $L_{b}$, the gratings
wash out quickly as a result of Doppler dephasing. The grating structure is
not lost, however. If the atoms interact with a second standing wave field,
the various spatial harmonics will rephase at different distances following
the interaction with the second field. By a proper choice of optical field
strengths, one can create and isolate atomic density patterns that
correspond very closely to pure, higher order spatial harmonics \cite{dubet1}%
. The patterns could be deposited on a substrate and used as elements in
soft x-ray systems.

One can also use nonresonant optical fields to interact with the atoms. In
this case, one creates a ground state phase grating rather than an amplitude
grating \cite{phase gratings}. If a highly-collimated atomic beam is used,
the phase grating can focus the atoms to a series of lines having spatial
period equal to $\lambda /2$. For a beam having angular divergence, echo
techniques can be used to rephase and isolate the various spatial harmonics 
\cite{dubet1,cahn}.

\begin{figure}[tb!]
\centering
\begin{minipage}{8.0cm}
\epsfxsize= 8 cm \epsfbox{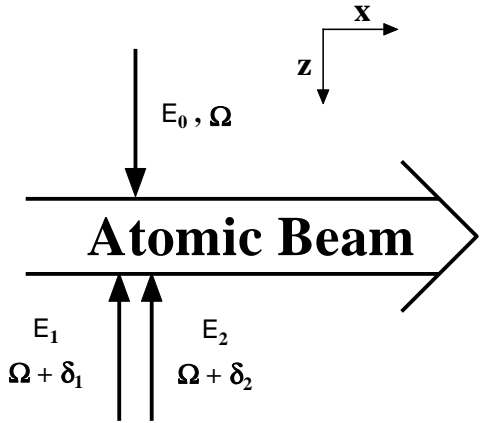}
\end{minipage}
\caption{Atom-field geometry. All fields overlap in the interaction region.
The detunings $\protect\delta _{1}$ and $\protect\delta _{2}$ are chosen
such that $\left| \protect\delta _{1}\right| T,\left| \protect\delta
_{2}\right| T\gg 1$, where $T$ is the interaction time in the atomic rest
frame.}
\label{fig1}
\end{figure}

\begin{figure}[tb!]
\centering
\begin{minipage}{8.0cm}
\epsfxsize= 8 cm \epsfbox{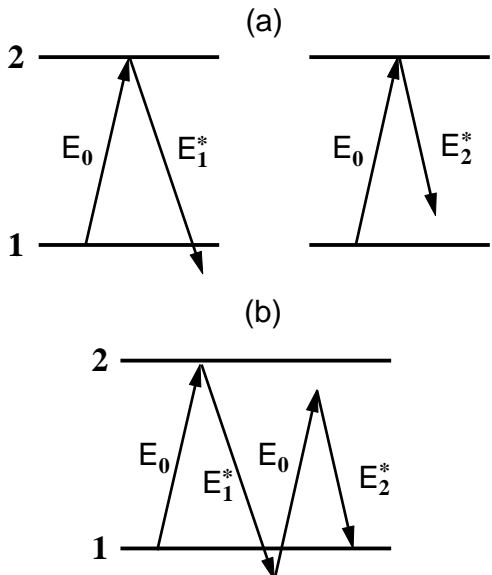}
\end{minipage}
\caption{Elementary processes involving fields $E_{0}$, $E_{1}$ and $E_{2}$.
(a) Two photon transitions involving field $E_{0}$ and either of fields $%
E_{1}$ or $E_{2}$ are not resonant (the diagrams are drawn with $\Omega
=\omega $). (b) A four photon transition is resonant provided that $\delta
_{1}=-\delta _{2}$.}
\label{fig2}
\end{figure}

If the goal of an experiment is to create a pure, higher order spatial
harmonic, it would be helpful to eliminate the lower order harmonics from
the outset. In this paper, we describe a method in which a single field
interaction zone can be used to produce gratings having spatial period equal
to $\lambda /(2n)$ where $n\geq 2$ (see Fig. \ref{fig1}). An atomic beam passes through an interaction region in which three fields
act. One of the fields has frequency $\Omega $ and propagates in the ${\bf 
\hat{z}}$ direction, while two additional fields, each counterpropagating
relative to the first field, have frequencies $\Omega +\delta _{1}$ and $%
\Omega +\delta _{2}$, respectively. A two-photon process in which a photon
of frequency $\Omega $ is absorbed and one of frequency $\Omega +\delta _{1}$
or $\Omega +\delta _{2}$ is emitted (see Fig. \ref{fig2}a) produces a contribution to
the ground state amplitude varying as $e^{2ikz}$, $k=2\pi /\lambda $. Such
two-photon processes are {\em not} resonant, however, and lead to a
vanishingly small second harmonic amplitude if $\left| \delta _{j}T\right|
\gg 1$ ($j=1,2)$, where $T$ is the pulse duration in the atomic rest frame.
On the other hand, an elementary {\em four-photon} process involving the
absorption of two photons of frequency $\Omega $ and the emission of one
each of frequency $\Omega +\delta _{1}$ and $\Omega +\delta _{2}$ {\em is}
resonant, provided $\delta _{1}=-\delta _{2}$ (see Fig. \ref{fig2}b). This process is
responsible for the creation of the fourth spatial harmonic in the ground
state amplitude. Thus the geometry of Fig. \ref{fig2}b can be used to produce a
ground state amplitude\ having spatial period $\lambda /4$ provided $\delta
_{1}=-\delta _{2}$. With other choices of $\delta _{1}$ and $\delta _{2},$
one can suppress an arbitrary number of lower order harmonics and produce a
ground state amplitude having period $\lambda /(2n)$ where $n>2$.

The suppression of lower order harmonics has important implications for
nanolithography. It is shown below that the field amplitudes and detunings
can be chosen such that a single, higher order atomic grating having period $%
\lambda /(2n)$ can be created to a high degree of accuracy. Both amplitude
and phase gratings can be produced. Phase gratings can lead to atom focusing
into lines separated by $\lambda /(2n)\ll \lambda /2$ for $n\gg 1$.
Amplitude gratings represent a pure, higher order spatial harmonic that can
be deposited on a substrate or used directly to scatter soft x-rays.

Of course, there are other methods for suppressing lower order spatial
harmonics. A coherent beam splitter can produce momenta separation of state
amplitudes that are greater than $2\hbar k$. As long as these momenta
components are not {\em spatially} separated, the atomic beam contains
density matrix elements in which lower order spatial harmonics are
suppressed. Beam splitters based on higher order Bragg scattering \cite
{Bragg} can lead to ground state densities corresponding very closely to
pure, higher order spatial harmonics. Adiabatic rapid passage \cite
{adiabatic rapid passage} can also lead to very nearly pure, higher order
spatial harmonics, albeit between state amplitudes corresponding to {\em %
different} internal states. Beam splitters based on triangular optical
potentials such as magneto-optical beam splitters \cite{magneto optical} and
bichromatic beam splitters \cite{bichromatic} produce large momenta
separations, but not pure spatial harmonics. Our approach is most closely
related to that of Yablonovich and Vrijen \cite{Yablonovich},\ who use
different frequency fields to increase spatial modulation in two-photon
microscopy. Here we extend and expand their ideas to the domain of atom
optics and atom interferometry.

The paper is organized as follows: In Section II, an illustrative example is
considered in which the second order spatial harmonic is suppressed. Methods
for creating both amplitude and phase gratings are discussed, as are the
effects of spontaneous decay during the pulse. Both numerical solutions and
an approximate analytical solution is considered in this section. In Sec.
III, numerical solutions are presented for the suppression of harmonics
beyond second order. It is shown that high contrast, amplitude gratings can
be created for arbitrary $n$. A summary of the results is given in Sec. IV,
and the implications for nanolithography are discussed.

\section{Illustrative Example}

Many of the features of harmonic suppression can be illustrated for the
field geometry of Fig. \ref{fig1}. The atomic beam propagates in the ${\bf \hat{x}}$
direction and the fields propagate along the z-axis. The total field can be
written as ${\bf E}({\bf r},t)={\bf E}(z,t)f(x,y),$ where $f(x,y)$ is a
spatial mode function, 
\begin{eqnarray}
{\bf E}(z,t) &=&\frac{1}{2}\widehat{{\bf y}}\left[ E_{0}e^{i(kz-\Omega
t)}+E_{1}e^{i[-kz-(\Omega +\delta _{1})t]}\right.  \nonumber \\
&&+\left. E_{2}e^{i[-kz-(\Omega +\delta _{2})t]}\right] +c.c.,  \label{1}
\end{eqnarray}
and $c.c$. stands for complex conjugate. Fields 1 and 2 propagate in the -$%
{\bf \hat{z}}$ direction and field 0 in the +${\bf \hat{z}}$ direction. In
this section, we set $E_{1}=E_{2}$, take $E_{1}$ and $E_{0}$ to be real, and
set $\delta _{1}=-\delta _{2}\equiv \delta $, such that 
\begin{equation}
{\bf E}(z,t)=\frac{1}{2}\widehat{{\bf y}}e^{-i\Omega t}\left[
E_{0}e^{ikz}+2E_{1}e^{-ikz}\cos (\delta t)\right] +c.c.  \label{2}
\end{equation}
In the atomic rest frame these fields appear as a radiation pulse and the
field amplitudes become functions of time. The atoms are modeled as having
two levels, 1 and 2, separated in frequency by $\omega .$ The problem
divides into two parts, interaction of the atoms with the fields and free
evolution of the atoms following the field interaction. In the main body of
the paper we consider only the field interaction region. In the field
interaction region, all effects associated with quantization of the atoms'
center-of-mass motion, as well as any transverse Doppler shifts, are
neglected. In Sec. IV, we will discuss the free evolution of the atoms
following their interaction with the field.

In the atomic rest frame, the state amplitudes $a_{1}$ and $a_{2}$ evolve as 
\begin{equation}
{\bf \dot{a}=}-i{\bf F(}z,t){\bf a}-2i{\bf G(}z,t{\bf )a},  \label{3}
\end{equation}
where 
\begin{eqnarray}
{\bf a} &=&\left( 
\begin{array}{c}
a_{1} \\ 
a_{2}
\end{array}
\right) ;  \nonumber \\
{\bf F(}z,t) &=&\left( 
\begin{array}{cc}
\Delta /2 & \chi _{0}(z,t)^{\ast } \\ 
\chi _{0}(z,t) & -\Delta /2
\end{array}
\right) ;  \nonumber \\
{\bf G}(z,t) &=&\left( 
\begin{array}{cc}
0 & \chi _{1}(z,t)^{\ast } \\ 
\chi _{1}(z,t) & 0
\end{array}
\right) \cos (\delta t),  \label{4}
\end{eqnarray}
$\chi _{0}(z,t)$ and $\chi _{1}(z,t)$ are Rabi frequencies defined by 
\begin{eqnarray}
\chi _{j}(z,t) &=&\chi _{j}(t)e^{ik_{j}z};  \nonumber \\
\chi _{j}(t) &=&-\mu E_{j}(t)/2\hbar =[\chi _{j}(t)]^{\ast }\geq 0;\text{ \{}%
j=0,1,2\};  \nonumber \\
k_{0} &=&-k_{1}=-k_{2}=k,  \label{5}
\end{eqnarray}
$\Delta =\Omega -\omega $ is an atom-field detuning, $\mu $ is a dipole
moment matrix element, and $E_{j}(t)$ is a pulse envelope function in the
atomic rest frame. Spontaneous decay during the pulse has been neglected for
the moment. Equation (\ref{3}) can be solved numerically for any pulse
envelopes $E_{j}(t)$. The advantage of the choice of parameters considered
in this section is that they allow for a very good approximate analytical
solution that illustrates the relevant physical concepts.

Since the two matrix elements of ${\bf G}$ have the same time dependence, it
is convenient to write a solution to Eq. (\ref{3}) in the form 
\begin{mathletters}
\label{6}
\begin{eqnarray}
{\bf a(}t) &=&{\bf S}(z,t)\,{\bf \tilde{a}(}t);  \label{6a} \\
{\bf S}(z,t) &=&\exp \left\{ -2i\int_{-\infty }^{t}{\bf G}(z,t^{\prime
})dt^{\prime }\right\}  \label{6b}
\end{eqnarray}
where the matrix ${\bf \tilde{a}(}t)$ satisfies the differential equation 
\end{mathletters}
\begin{equation}
d{\bf \tilde{a}/}dt=-i{\bf S}^{\dagger }{\bf (}z,t){\bf F(}z,t{\bf )S(}z,t%
{\bf )\tilde{a}.}  \label{7}
\end{equation}
For a smoothly varying pulse having duration $T\gg $ $\delta ^{-1}$, the
integral in Eq. (\ref{6b}) can be evaluated asymptotically. It follows that 
\begin{eqnarray}
{\bf S(}z,t{\bf )} &\sim &{\bf \exp }\left\{ -2i\frac{\sin (\delta t)}{%
\delta }\left( 
\begin{array}{cc}
0 & \chi _{1}(z,t)^{\ast } \\ 
\chi _{1}(z,t) & 0
\end{array}
\right) \right\}  \nonumber \\
&=&\cos (\theta )\,{\bf 1}-i\,\sin (\theta )\left( 
\begin{array}{cc}
0 & e^{ikz} \\ 
e^{-ikz} & 0
\end{array}
\right) ,  \label{8}
\end{eqnarray}
where 
\begin{equation}
\theta =2\chi _{1}(t)\sin (\delta t)/\delta ,  \label{9}
\end{equation}
{\bf 1} is the unit matrix, and we have used the fact that $\chi
_{1}(z,t)/\left| \chi _{1}(z,t)\right| =\exp (-ikz)$. The S-matrix reduces
to the unit matrix as $t\sim \pm \infty .$ Combining Eqs. (\ref{6})- (\ref{9}%
), one finds 
\begin{eqnarray}
d{\bf \tilde{a}/}dt &=&-i\frac{\Delta }{2}{\bf \cos (}2\theta )\left( 
\begin{array}{cc}
1 & 0 \\ 
0 & -1
\end{array}
\right) {\bf \tilde{a}}-i\left( 
\begin{array}{cc}
0 & M \\ 
M^{\ast } & 0
\end{array}
\right) {\bf \tilde{a}}  \label{10} \\
&&{\bf +}\frac{\sin (2\theta )}{2}\left( 
\begin{array}{cc}
N & -\Delta e^{ikz} \\ 
\Delta e^{-ikz} & N^{\ast }
\end{array}
\right) {\bf \tilde{a}}
\end{eqnarray}
where 
\begin{mathletters}
\label{11}
\begin{eqnarray}
M &=&\chi _{0}(t)\left[ \cos ^{2}(\theta )e^{-ikz}+\sin ^{2}(\theta )e^{3ikz}%
\right] ;  \label{11a} \\
N &=&2i\chi _{0}(t)\sin (2kz).  \label{11b}
\end{eqnarray}

Equation (\ref{10}) can be simplified considerably if 
\end{mathletters}
\begin{equation}
\delta \gg \chi _{0}(t),\text{ }\Delta ,\text{ }\chi _{0}(t)\chi
_{1}^{2}(t)/\delta ^{2}.  \label{11p}
\end{equation}
In that limit, it is possible to ''course-grain'' Eq. (\ref{10}) on a time
scale greater than $\delta ^{-1}$ and replace all trigonometric functions
appearing in Eqs. (\ref{10}) and (\ref{11}) by their time averages. Using
Bessel function expansions for the trigonometric functions, one finds 
\begin{eqnarray}
\alpha &\equiv &\overline{\cos ^{2}(\theta )}=\frac{1+J_{0}[4\chi
_{1}(t)/\delta ]}{2};  \nonumber \\
\beta &\equiv &\overline{\sin ^{2}(\theta )}=\frac{1-J_{0}[4\chi
_{1}(t)/\delta ]}{2};\qquad \overline{\sin (2\theta )}=0,  \label{12}
\end{eqnarray}
where $J_{0}$ is a Bessel function and the bar indicates a time average. In
this limit, Eq. (\ref{10}) is replaced by 
\begin{eqnarray}
d{\bf \tilde{a}/}dt &=&-i\frac{\Delta }{2}J_{0}[4\chi _{1}(t)/\delta ]\left( 
\begin{array}{cc}
1 & 0 \\ 
0 & -1
\end{array}
\right) {\bf \tilde{a}}  \nonumber \\
&&-i\chi _{0}(t)\left( 
\begin{array}{cc}
0 & \alpha e^{-ikz}+\beta e^{3ikz} \\ 
\alpha e^{ikz}+\beta e^{-3ikz} & 0
\end{array}
\right) {\bf \tilde{a}.}  \label{13}
\end{eqnarray}
Equation (\ref{13}) admits solutions which represent both amplitude and
phase gratings in the ground state atomic density. We examine these cases
separately.

\subsection{Amplitude gratings\qquad $\Delta =0$}

It is possible to obtain amplitude gratings having maximum contrast for the
ground and excited state populations by choosing $\Delta =0$ ($\Omega
=\omega ).$ In this limit, Eq. (\ref{13}) reduces to 
\begin{equation}
d{\bf \tilde{a}/}dt=-i\chi _{0}(t)\left( 
\begin{array}{cc}
0 & \alpha e^{-ikz}+\beta e^{3ikz} \\ 
\alpha e^{ikz}+\beta e^{-3ikz} & 0
\end{array}
\right) {\bf \tilde{a}.}  \label{14}
\end{equation}

\begin{figure}[tb!]
\centering
\begin{minipage}{8.0cm}
\epsfxsize= 8 cm \epsfysize= 2.5 cm \epsfbox{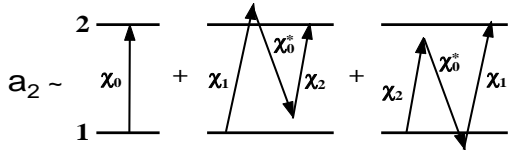}
\end{minipage}
\caption{Elementary processes leading to the resonant excitation of state 2.
The Rabi frequencies $\chi _{j}$ are a shorthand notation for $\chi
_{j}(z,t) $ and contain spatial phase factors. In forming the upper state
probability, interference terms lead to a spatial modulation having period $%
\lambda /4$.}
\label{fig3}
\end{figure}

It will prove useful to look at the perturbative limit of Eq. (\ref{14}).
When $\chi _{1}(t)/\delta \ll 1,$ the time evolution of state 2 is 
\begin{equation}
d\tilde{a}_{2}{\bf /}dt=-i\chi _{0}(t)e^{ikz}\tilde{a}_{1}-2i\frac{\chi
_{0}(t)[\chi _{1}(t)]^{2}}{\delta ^{2}}e^{-3ikz}\tilde{a}_{1}.  \label{15}
\end{equation}
The fundamental processes responsible for excitation to state 2 from state 1
are shown schematically in Fig. \ref{fig3}. There can be direct, resonant excitation
to state 2 by field $E_{0}$ and {\em three-photon}, resonant excitation
involving the absorption of one photon each from fields $E_{1}$ and $E_{2}$
and emission of one photon into field $E_{0}$ (as noted above, the amplitude
modulated field used in this section can be considered as a sum of two
fields having frequencies $\Omega _{1}=\Omega +\delta _{1}$ and $\Omega
_{2}=\Omega +\delta _{2}$, with $\delta _{1}=-\delta _{2}\equiv \delta $).
The three-photon processes are resonant since $\Omega _{1}+$ $\Omega
_{2}-\Omega =\omega $ when $\Omega =\omega $. In these diagrams, the Rabi
frequency $\chi _{0}$ is a shorthand notation for $\chi _{0}(z,t)$ and
contains a factor $e^{ikz}$ and, similarly, Rabi frequencies $\chi _{1}$ and 
$\chi _{2}$ contain factors $e^{-ikz}$. The overall amplitude for the one
photon process varies as $\chi _{0}e^{ikz}$ and as ($\chi _{0}\chi _{1}\chi
_{2})e^{-3ikz}/\delta ^{2}$ for the three-photon processes. The $\delta ^{2}$
factor reflects the fact that the two intermediate states in the
three-photon processes are each off-resonance by an energy $\hbar \delta $.
In taking the square of the amplitude, one finds a spatial modulation at $%
4kz.$ As seen in Sec. III, diagrams of this type can be used to estimate the
required field strengths for suppression of higher order harmonics.

\begin{figure}[tb!]
\centering
\begin{minipage}{8.0cm}
\epsfxsize= 8 cm \epsfysize= 8 cm \epsfbox{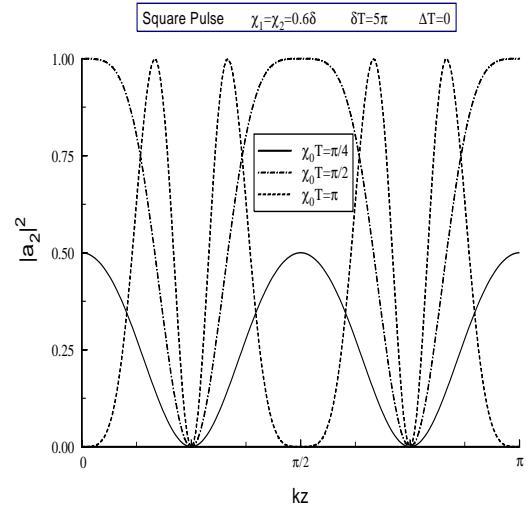}
\end{minipage}
\caption{Graphs of the upper state probability as a function of $kz$ for
excitation by a rectangular pulse, with $\delta _{1}=-\delta _{2}\equiv
\delta $.}
\label{fig4}
\end{figure}

Equation (\ref{14}) cannot be solved analytically for arbitrary pulse
envelope functions. On the other hand, it {\em can} be solved analytically
if we assume square pulses, $E_{j}(t)=E_{j}$ for $0\leq t\leq T$ and zero
otherwise. [Equation (\ref{9}) remains valid for a square pulse. Moreover,
if one chooses a detuning $\delta T=n\pi $, then the S-matrix in Eq. (\ref{8}%
) reduces to the unit matrix at the beginning and end of the pulse. Thus one
can use Eq. (\ref{14}), even though a square pulse does not satisfy the
adiabaticity conditions at turn on and turn off.] For a square pulse, one
finds 
\begin{eqnarray}
\left| a_{2}(T)\right| ^{2} &=&\left| \tilde{a}_{2}(T)\right| ^{2}=\sin ^{2}%
\left[ \chi _{0}T\left| \alpha e^{-ikz}+\beta e^{3ikz}\right| \right] 
\nonumber \\
&=&\sin ^{2}\left\{ \chi _{0}T\left( \left[ (1+J_{0}^{2}\{4\chi _{1}/\delta
\})\right. \right. \right.  \nonumber \\
&&\left. \left. \left. +(1-J_{0}^{2}\{4\chi _{1}/\delta \})\cos 4kz\right]
/2\right) ^{1/2}\right\} .  \label{16}
\end{eqnarray}
It follows from Eq. (\ref{16}) [or from Eq. (\ref{14})] that the excited
state population immediately after the pulse is a periodic function of $z$
having period $\lambda /4$. The second order spatial harmonic having period $%
\lambda /2$ has been $\sup $pressed, owing to the large detunings $\pm
\delta $ of the fields propagating in the $-{\bf \hat{z}}$ direction. By
taking a field strength corresponding to the first zero of the Bessel
function $J_{0}[4\chi _{1}/\delta ]$ ($\chi _{1}\sim 0.6\delta )$, one can
optimize the grating contrast for the smallest possible value of $\chi _{0}T$%
. If $J_{0}[4\chi _{1}/\delta ]=0,$ then $\left| a_{2}(T)\right| ^{2}=\sin
^{2}\left[ \chi _{0}T\cos (2kz)\right] $. This function is plotted as a
function of $kz$ in Fig. \ref{fig4} for several values of $\chi _{0}T$. 

For values of 
$\chi _{0}T\ll \pi /2$, the grating contrast in the excited state amplitude
is less than unity, but the grating is close to pure fourth harmonic. For $%
\chi _{0}T\geq \pi /2$, the grating contrast is always unity, but harmonics
higher than fourth are evident. A choice of pulse area that leads to high
grating contrast while producing a spatial distribution very close to pure
fourth harmonic is $\chi _{0}T=1.3$, for which the Bessel function expansion
of $\sin ^{2}\left[ \chi _{0}T\cos (2kz)\right] $ yields 
\begin{eqnarray}
\left| a_{2}(T)\right| ^{2} &\sim &\left[ 1-J_{0}(2.6)+J_{2}(2.6)\cos
(4kz)\right.  \nonumber \\
&&\left. -J_{4}(2.6)\cos (8kz)+\ldots \right] /2  \nonumber \\
&\approx &[1.1+0.92\cos (4kz)-0.17\cos (8kz)]/2.  \label{17}
\end{eqnarray}
This grating has a fringe contrast of 0.93 and consists primarily of fourth
harmonic.

The qualitative results do not change when smooth pulses are used. The
original differential equation (\ref{4}) was solved numerically for Gaussian
pulses, 
\begin{equation}
\chi _{j}(t)=\chi _{j}e^{-(t/T)^{2}}/\sqrt{\pi }.  \label{18}
\end{equation}
From this point onward, we set $T=1$ and evaluate all times in units of $T$
and all frequencies in units of $T^{-1}$. In these units, the quantity $%
2\chi _{j}$ corresponds to the area of pulse $j$. In all the examples, $%
\delta \geq 3$, which, for the Gaussian pulse (\ref{18}), is sufficient to
guarantee adiabaticity, provided\ that conditions (\ref{11p}) are also met.

\begin{figure}[tb!]
\centering
\begin{minipage}{8.0cm}
\epsfxsize= 8 cm \epsfysize= 8 cm \epsfbox{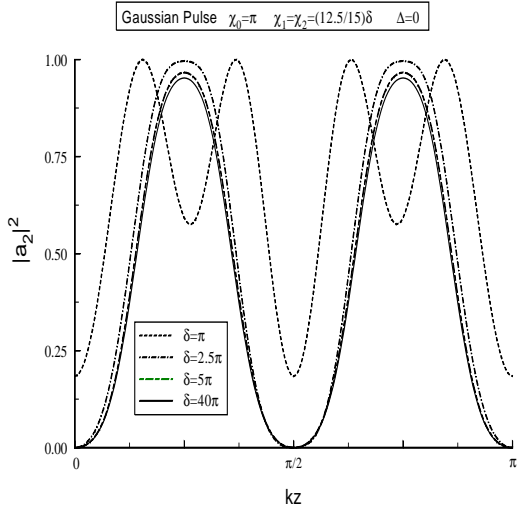}
\end{minipage}
\caption{Graphs of the upper state probability as a function of $kz$ for
excitation by a Gaussian pulse, with $\delta _{1}=-\delta _{2}\equiv \delta $%
. In this and subsequent graphs, all frequencies are in units of $T^{-1}$.
With increasing ratio of $\delta /\chi _{0}$, the solution of the exact
equations (\ref{4}) approaches that of the course-grained equations (\ref{15}%
), represented by the solid line in the plot.}
\label{fig5}
\end{figure}

In Fig. \ref{fig5}, the solutions of the exact equations [Eq. (\ref{4})] for the
upper state probability are compared with those of the coarse-grained
equations [Eq. (\ref{14})]. Recall that the solutions of the course grained
equations should approach those of the exact equations as the ratio $\delta
/\chi _{0}$ increases. This feature is seen clearly in Fig. 5. The curve
having $\delta /\chi _{0}=40$ correspond to the solution of both the exact
and course-grained equations - they are not distinguishable for this ratio
of $\delta /\chi _{0}.$ Even for $\delta /\chi _{0}=5,$ the results do not
differ by much.

\begin{figure}[tb!]
\centering
\begin{minipage}{8.0cm}
\epsfxsize= 8 cm \epsfysize= 8 cm \epsfbox{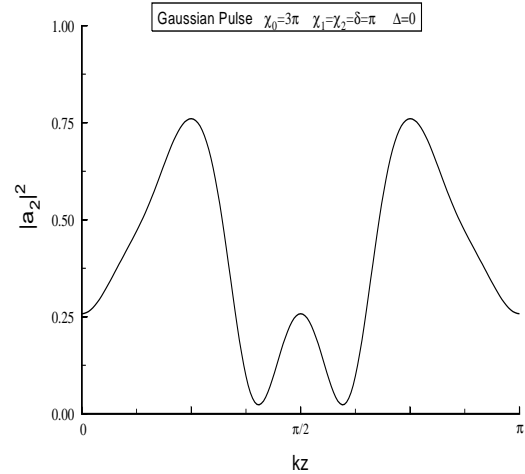}
\end{minipage}
\caption{Graphs of the upper state probability as a function of $kz$ for
excitation by a Gaussian pulse, with $\delta _{1}=-\delta _{2}\equiv \delta $
and $\delta /\chi _{0}=1/3$. For this ratio, the ''course-graining''
approximation is no longer valid and the period of the amplitude grating is $%
\lambda /2$ rather than $\lambda /4$.}
\label{fig6}
\end{figure}

The course grained equations depend only on the parameters $%
\chi _{0}$ and $\delta /\chi _{1}$; consequently the solutions of the exact
equations depend on these two parameters only for $\delta /\chi _{0}\gg 1$.
If $\delta /\chi _{0}\lesssim 1$, the solution of the exact equations
depends independently on the parameters $\chi _{0},$ $\delta $, and $\chi
_{1}$; moreover, the solution has period $\lambda /2$ rather than $\lambda
/4 $. Although not evident from the figure, the $\delta =\chi _{0}$ results
differ from pure $\lambda /4$ periodicity by as much as 10\%. These
differences are more evident in Fig. \ref{fig6}, drawn for $\delta /\chi _{0}=1/3$.
To assure $\lambda /4$ periodicity, one must have $\delta /\chi _{0}\gtrsim
5 $.
\begin{figure}[tb!]
\centering
\begin{minipage}{8.0cm}
\epsfxsize= 8 cm \epsfysize= 8 cm \epsfbox{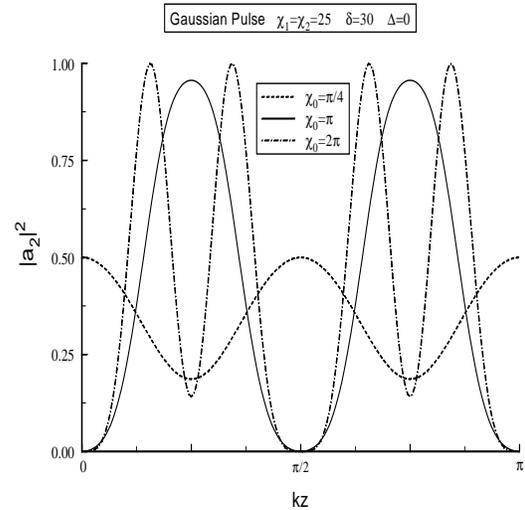}
\end{minipage}
\caption{Graphs of the upper state probability as a function of $kz$ for
excitation by a Gaussian pulse, with $\delta _{1}=-\delta _{2}\equiv \delta $
and several values of $\chi _{0}$.}
\label{fig7}
\end{figure}

\begin{figure}[tb!]
\centering
\begin{minipage}{8.0cm}
\epsfxsize= 8 cm \epsfysize= 8 cm \epsfbox{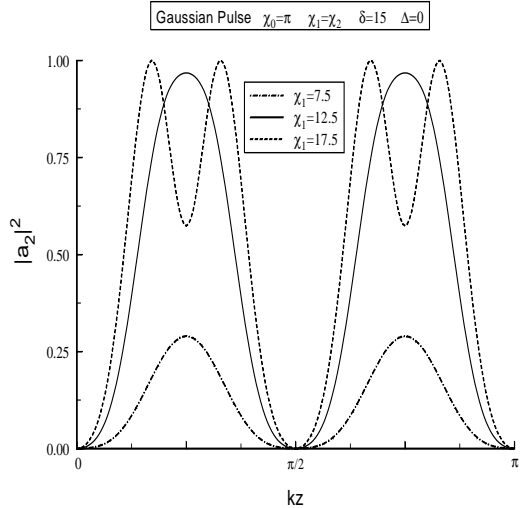}
\end{minipage}
\caption{Graphs of the upper state probability as a function of $kz$ for
excitation by a Gaussian pulse, with $\delta _{1}=-\delta _{2}\equiv \delta $
and several values of $\chi _{1}$.}
\label{fig8}
\end{figure}

In Fig. \ref{fig7}, the upper state probability is plotted as a function of $kz$ for
fixed $\chi _{1}/\delta =5/6$, and several values of $\chi _{0}$. These
curves follow the same qualitative behavior as those shown in Fig. \ref{fig4} for the
square pulse. In Fig. \ref{fig8}, the upper state probability is plotted as a
function of $kz$ for fixed $\chi _{0}=\pi $ and several values of $\chi
_{1}/\delta $. For $\chi _{1}/\delta \lesssim 0.5$, the grating contrast is
less than unity, but the grating is very nearly pure fourth harmonic. With
increasing $\chi _{1}/\delta $, the grating contrast approaches unity and
harmonics of order ($4+4n)$, $n\geq 1$, are evident.

\subsection{Phase Gratings\qquad $\Delta >1$}

\begin{figure}[tb!]
\centering
\begin{minipage}{8.0cm}
\epsfxsize= 8 cm \epsfysize= 3.2 cm \epsfbox{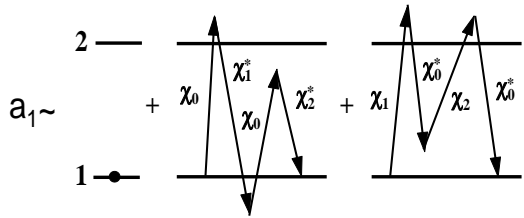}
\end{minipage}
\caption{Elementary processes leading to a phase grating in the ground state
amplitude. The second and third diagrams in the figure are spatially
modulated with period $\lambda /4$.}
\label{fig9}
\end{figure}

When $\Delta >1$, the elementary processes shown in Fig. \ref{fig3} are no longer
resonant. Thus one expects that the excited state population following the
pulse to be negligible. On the other hand the resonant, elementary 4-photon
processes, shown in Fig. \ref{fig9}, lead to a modification of the {\em phase} of the
ground state amplitude.

\begin{figure}[tb!]
\centering
\begin{minipage}{8.0cm}
\epsfxsize= 8 cm \epsfysize= 8 cm \epsfbox{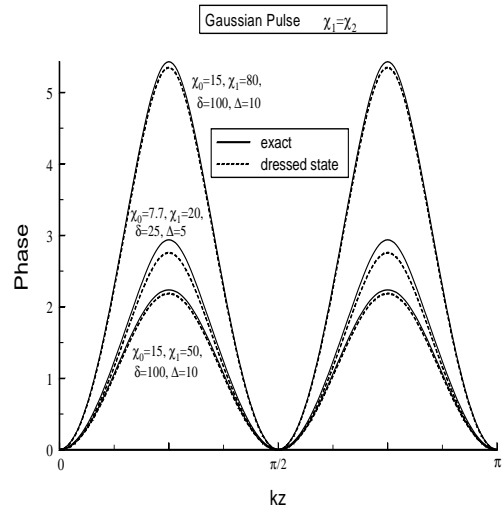}
\end{minipage}
\caption{Graphs of the ground state phase as a function of $kz$ for
excitation by a Gaussian pulse, with $\delta _{1}=-\delta _{2}\equiv \delta $%
. The solid lines are a solution of the exact equations while the dashed
lines are obtained from the course-grained, dressed state equations.}
\label{fig10}
\end{figure}

To calculate the phase change of the ground state
amplitude, it is convenient to use the semiclassical dressed states
associated with Eq. (\ref{14}). These states are obtained by instantaneously
diagonalizing Eq. (\ref{14}). If 
\begin{eqnarray}
&&\Delta ^{2}J_{0}^{2}\left\{ \frac{4\chi _{1}(t)}{\delta }\right\} 
\nonumber \\
&&+2\chi _{0}(t)^{2}\left\{ \left( 1+J_{0}^{2}\left\{ \frac{4\chi _{1}(t)}{%
\delta }\right\} \right) \right.  \nonumber \\
&&+\left. \left( 1-J_{0}^{2}\left\{ \frac{4\chi _{1}(t)}{\delta }\right\}
\right) \cos 4kz\right\} \left. \gg 1\right.  \label{18a}
\end{eqnarray}
the system remains in the instantaneous eigenstate that evolves from the
ground state as the field is turned on and returns to the ground state
following the pulse. The net phase change is simply the integral of the
dressed state energy divided by $\hbar .$ Explicitly, one finds. 
\begin{equation}
a_{1}(\infty )=e^{-i\phi },  \label{19}
\end{equation}
where 
\begin{eqnarray}
\phi &=&\frac{1}{2}\int_{-\infty }^{\infty }dt\left\{ \Delta
^{2}J_{0}^{2}\left\{ \frac{4\chi _{1}(t)}{\delta }\right\} \right.  \nonumber
\\
&&+2\chi _{0}(t)^{2}\left[ \left( 1+J_{0}^{2}\left\{ \frac{4\chi _{1}(t)}{%
\delta }\right\} \right) \right.  \nonumber \\
&&+\left. \left. \left( 1-J_{0}^{2}\left\{ \frac{4\chi _{1}(t)}{\delta }%
\right\} \right) \cos 4kz\right] \right\} ^{1/2}  \label{19a}
\end{eqnarray}
For Gaussian pulses (\ref{18}), the phase $\phi $ is plotted as a function
of $kz$ in Fig. \ref{fig10} for several values of $\chi _{0},$ $\chi _{1}$, $\delta $%
, and $\Delta $.

The solid line represents solutions of the exact equations (%
\ref{4}) and the dotted line the values given by Eq. (\ref{19a}), with the
phase arbitrarily taken equal to zero at $kz=0$. The results agree when
conditions (\ref{18a}) and (\ref{11p}) are satisfied. For $\chi _{1}/\delta
\geq 0.6$, there are times during the atom-field interaction for which $%
J_{0}\left\{ \frac{4\chi _{1}(t)}{\delta }\right\} =0.$ As a consequence,
condition (\ref{18a}) is violated near $kz=\pi /4$ and $3\pi /4$. This
explains the small deviations between the exact and approximate solutions
for these values of $kz$. For all cases shown, $\left| a_{1}(\infty )\right|
\sim 1$. It is seen that significant phase modulation of the ground state
can be achieved with suppression of the second spatial harmonic.

A necessary condition for significant spatial modulation of the phase is $%
\chi _{0}\chi _{1}^{2}/\delta ^{2}\geq 1$. If $\Delta ^{2}J_{0}^{2}\left\{ 
\frac{4\chi _{1}(t)}{\delta }\right\} \ll \chi _{0}(t)^{2}$, then the
effective pulse area for the creation of the phase grating is of order $\chi
_{0},$ while it is of order $\chi _{0}^{2}/\Delta $ for $\Delta
^{2}J_{0}^{2}\left\{ \frac{4\chi _{1}(t)}{\delta }\right\} \gg \chi
_{0}(t)^{2}$.

\subsection{Role of Spontaneous Emission}

So far we have neglected the role of spontaneous emission during the
excitation pulse. To investigate the role of spontaneous emission during the
pulse, we adopt a {\em highly simplified} decay scheme in which state 2
decays to a state outside the two-state manifold with rate 2$\gamma $. As a
result, population leaks out of the two-level manifold. Clearly, any excited
state amplitude grating will be diminished by decay. It also turns out that
phase gratings are destroyed by spontaneous decay since the excited state
population {\em during} the pulse is not negligible under conditions
favoring a significant spatially modulated phase grating. Spontaneous
emission breaks the adiabatic following that would return the atom to its
ground state in the absence of such decay \cite{Robinson}.

Within the context of this model, spontaneous emission can be included in
Eq. (\ref{4}) by the addition of a term 
\[
-\gamma \left( 
\begin{array}{cc}
0 & 0 \\ 
0 & 1
\end{array}
\right) {\bf a}. 
\]
This decay term results in an addition to Eq. (\ref{14}) of the form 
\[
-\gamma \left( 
\begin{array}{cc}
\frac{1-J_{0}[4\chi _{1}(t)/\delta ]}{2} & 0 \\ 
0 & \frac{1+J_{0}[4\chi _{1}(t)/\delta ]}{2}
\end{array}
\right) {\bf \tilde{a}}. 
\]
For any nonvanishing field strength $\chi _{1}$, {\em both} dressed state
amplitudes $\tilde{a}_{1}$ and $\tilde{a}_{2}$ decay. For the field strength
giving rise to maximum spatial modulation, $J_{0}[4\chi _{1}(t)/\delta ]\sim
0$, spontaneous emission destroys grating formation when $\gamma \gtrsim 1$.
This is easily understood. To have strong spatial modulation of amplitude or
phase gratings, there is always a significant excited state population {\em %
during }the pulse; any decay results in a loss of population from the system
in our simple decay scheme.

We have also carried out density matrix calculations using various decay
schemes. It appears that one must require $\gamma \lesssim 0.5$ to retain
reasonable grating contrast.

\section{Suppression of Higher Order Harmonics}

One can generalize the previous results to allow for the suppression of
harmonics higher than second order. The appropriate field geometry is shown
in Fig. \ref{fig1}. Spontaneous emission during the atom-pulse interaction is
neglected. For the field (\ref{1}), the state amplitudes evolve as 
\begin{equation}
{\bf \dot{a}=}-i{\bf F(}z,t){\bf a}-i{\bf H(}z,t{\bf )a}  \label{20}
\end{equation}
where ${\bf F(}z,t)$ is defined by (\ref{4}) and 
\begin{equation}
{\bf H}(z,t)=\left( 
\begin{array}{cc}
0 & 
\begin{array}{c}
\chi _{1}(z,t)^{\ast }e^{i\delta _{1}t} \\ 
+\chi _{2}(z,t)^{\ast }e^{i\delta _{2}t}
\end{array}
\\ 
\begin{array}{c}
\chi _{1}(z,t)e^{-i\delta _{1}t} \\ 
+\chi _{2}(z,t)e^{-i\delta _{2}t}
\end{array}
& 0
\end{array}
\right) .  \label{21}
\end{equation}
Since ${\bf H}(z,t)$ cannot be factored as a constant matrix times a
function of $t$, the method used in Sec. II is not applicable. On the other
hand it is not difficult to obtain numerical solutions of Eq. (\ref{21}).

Diagrams similar to those shown in Figs. \ref{fig3} and \ref{fig9} can help us to estimate the
field strengths needed to obtain both amplitude and phase gratings. Before
considering such diagrams, we note that the lowest order ac Stark shift, 
\begin{equation}
\xi =\left( \chi _{1}^{2}/\delta _{1}+\chi _{2}^{2}/\delta _{2}\right)
\label{21a}
\end{equation}
no longer vanishes automatically as it does for an amplitude modulated field
($\chi _{1}=\chi _{2};\quad \delta _{1}=-\delta _{2})$. To suppress the
higher order harmonics, it is necessary that $\left| \delta _{1}\right| $
and $\left| \delta _{2}\right| $ be large compared with all characteristic
frequencies in the problems, including the ac Stark shifts. In order to have 
$\left| \delta _{1}\right| ,\left| \delta _{2}\right| >\xi $, it is
necessary that $\chi _{1}^{2}/\delta _{1}^{2}\ll 1$, $\chi _{2}^{2}/\delta
_{2}^{2}\ll 1$, {\em unless} $\delta _{1}$ and $\delta _{2}$ have opposite
signs and the field strengths are chosen such that 
\begin{equation}
\left( \chi _{1}^{2}/\delta _{1}+\chi _{2}^{2}/\delta _{2}\right) =0.
\label{22}
\end{equation}
In this case it is possible to have $\chi _{1}\sim \delta _{1}$ and still
have $\left| \delta _{1}\right| ,\left| \delta _{2}\right| >\xi $. Since it
is necessary to have $\chi _{1}\sim \delta _{1}$ for significant spatial
modulation in the state populations, it is important to choose the Rabi
frequencies such that Eq. (\ref{22}) is satisfied. Higher order
contributions to the ac Stark effect may still play a role. From the
numerical solutions, it appears that one must choose $\chi _{1}$ and $\chi
_{2}$ less than or of order $\left| \delta _{1}\right| $,$\left| \delta
_{2}\right| ,$ to obtain the desired periodicity.

\begin{figure}[tb!]
\centering
\begin{minipage}{8.0cm}
\epsfxsize= 8 cm \epsfysize= 3 cm \epsfbox{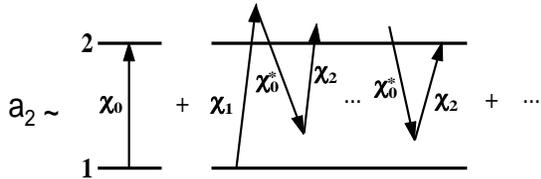}
\end{minipage}
\caption{Elementary processes leading to the resonant excitation of state 2
and suppression of spatial harmonics higher than fourth. The detunings are
chosen such that $n_{1}\delta _{1}+n_{2}\delta _{2}=0,$ leading to an
amplitude grating having period $\lambda /[2(n_{1}+n_{2})]$.}
\label{fig11}
\end{figure}

The elementary processes giving rise to an amplitude grating are illustrated
in Fig. \ref{fig11}. The grating is formed by terms representing the interference of
the two diagrams shown in the figure.

The resonance condition for the second
diagram is 
\begin{equation}
n_{1}\delta _{1}+n_{2}\delta _{2}=0  \label{23}
\end{equation}
when $\Delta =0$ and the integers $n_{1}>$ $n_{2}$ contain no common
factors. The lowest order nonvanishing harmonic varies as cos[2($%
n_{1}+n_{2})kz]$. The amplitude associated with this harmonic is of order 
\begin{equation}
A(n_{1},n_{2})=\frac{\chi _{0}^{n_{1}+n_{2}}\chi _{1}^{n_{1}}\chi
_{2}^{n_{2}}}{\delta ^{2(n_{1}+n_{2}-1)}},  \label{24}
\end{equation}
where $\delta $ is of order $\left| \delta _{1}\right| $ or $\left| \delta
_{2}\right| $. To obtain good grating contrast, one requires that 
\begin{equation}
A(n_{1},n_{2})\sim 1\text{.}  \label{25}
\end{equation}
\begin{figure}[tb!]
\centering
\begin{minipage}{8.0cm}
\epsfxsize= 8 cm \epsfysize= 8 cm \epsfbox{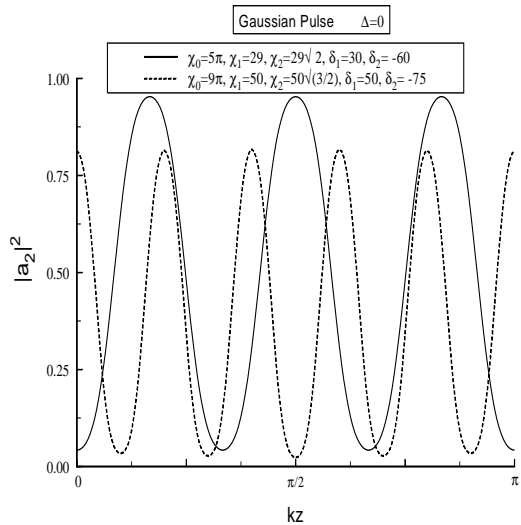}
\end{minipage}
\caption{Graphs of the upper state probability as a function of $kz$ for
excitation by a Gaussian pulse, with $n_{1}\delta _{1}+n_{2}\delta _{2}=0.$
Graphs are shown for ($n_{1}=2$, $n_{2}=1$) (sixth harmonic) and ($n_{1}=3$, 
$n_{2}=2$) (tenth harmonic).}
\label{fig12}
\end{figure}

Condition (\ref{25}) can be satisfied if $\chi _{0}\sim \delta
^{1-2/(n_{1}+n_{2})}$, but, to suppress lower harmonics, it is also
necessary that $\chi _{0}\ll \delta $. As ($n_{1}+n_{2})$ increases, $\delta 
$ and $\chi _{0}$ must be taken larger and larger to simultaneously satisfy
these conditions.

In Fig. \ref{fig12} the upper state population is plotted as a function of $kz$ for ($%
n_{1}=2$, $n_{2}=1$) (sixth harmonic) and for ($n_{1}=3$, $n_{2}=2$) (tenth
harmonic). The field strengths are chosen such that $\chi _{2}=\sqrt{%
n_{1}/n_{2}}\chi _{1}$ according to Eqs. (\ref{22}) and (\ref{23}), and $%
\chi _{1}$ is taken to be of order $\delta _{1}$. It is seen that amplitude
gratings having periods as small as $\lambda /10$ can be produced with
contrast of order unity. For sufficiently large $\delta $, an arbitrary
number of lower order harmonics can be suppressed.
\begin{figure}[tb!]
\centering
\begin{minipage}{8.0cm}
\epsfxsize= 8 cm \epsfysize= 3 cm \epsfbox{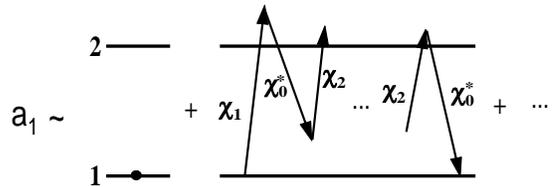}
\end{minipage}
\caption{Elementary processes leading to a ground state phase grating and
suppression of spatial harmonics higher than fourth. The detunings are
chosen such that. $n_{1}\delta _{1}+n_{2}\delta _{2}=0,$ leading to a phase
grating having period $\lambda /[2(n_{1}+n_{2})]$.}
\label{fig13}
\end{figure}

\begin{figure}[tb!]
\centering
\begin{minipage}{8.0cm}
\epsfxsize= 8 cm \epsfysize= 8 cm \epsfbox{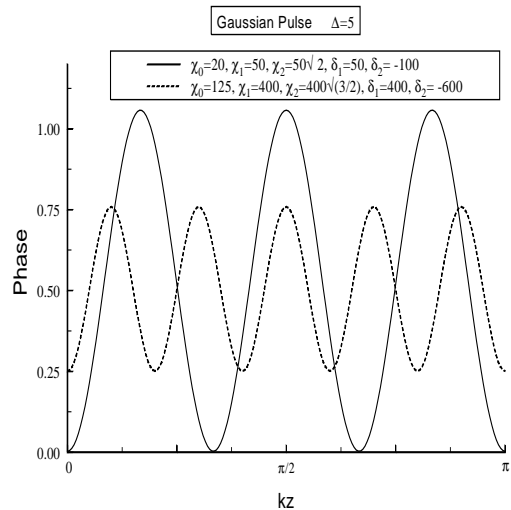}
\end{minipage}
\caption{Graphs of the ground state phase as a function of $kz$ for
excitation by a Gaussian pulse, with $n_{1}\delta _{1}+n_{2}\delta _{2}=0.$
Graphs are shown for ($n_{1}=2$, $n_{2}=1$) (sixth harmonic) and ($n_{1}=3$, 
$n_{2}=2$) (tenth harmonic).}
\label{fig14}
\end{figure}

The conditions for producing phase gratings are similar, but a little more
severe, than those for producing amplitude gratings, provided one takes $%
\Delta >1$, but $\chi _{0}\gg \Delta .$ The elementary processes giving rise
to an phase grating are illustrated in Fig. \ref{fig13}. The highest contrast phase
gratings are produced when $A(n_{1},n_{2})\sim \Delta $, which can be
satisfied if $\chi _{0}\sim \delta ^{1-2/(n_{1}+n_{2})}\Delta
^{1/(n_{1}+n_{2})}$. Along with the requirement that $\chi _{0}\ll \delta $,
this condition implies that large values of $\delta $ and $\chi _{0}$ are
needed to produce large phase gratings as ($n_{1}+n_{2})$ increases. The
phase of the ground state amplitude is shown as a function of $kz$ in Fig.
\ref{fig14} for ($n_{1}=2$, $n_{1}=1$) (sixth harmonic) and for ($n_{1}=3$, $n_{2}=2$%
) (tenth harmonic). In all cases shown, $\left| a_{1}(\infty )\right| \sim 1$
. Clearly it is also possible to produce phase gratings having reasonable
contrast with the suppression of lower order harmonics.

\section{Discussion}

It has been shown that atom-field interactions can be used to create high
contrast amplitude gratings for atomic state populations and phase gratings
for atomic ground state amplitudes. If counterpropagating fields having
wavelength $\lambda $ are used, it is possible to choose atom-field
detunings such that the lowest spatial harmonic component in the gratings
has period $\lambda /2n$, where $n$ is an integer. For the sake of
definiteness, we will refer to the harmonic having period $\lambda /2n$ as
the (2$n)$th harmonic. The gratings produced are not pure in the sense that
they contain all integer multiples of the (2$n)$th harmonic, but the field
strengths can be chosen in such a manner as to maximize the contribution
from the (2$n)$th harmonic. Once the gratings are created, the question
remains as how to image the gratings at some distance $L$ from the
atom-field interaction zone. One can probe the gratings by applying
counterpropagating fields having frequencies $\Omega $ and $\Omega +\delta
_{1}$ to generate a signal with frequency $\Omega +\delta _{2}$;
alternatively, one can deposit the atomic density grating on a substrate.
The situation is different for amplitude and phase gratings.

In the case of amplitude gratings, one has the option of working with
distances $L$ that correspond to either classical or quantum-mechanical
scattering for the atoms. If $L\gtrsim L_{T},$ where the so-called Talbot
length $L_{T}$ is defined by $L_{T}=2(\lambda /2n)^{2}/\lambda _{dB}$ ($%
\lambda _{dB}$ is the atomic de Broglie wavelength), the atomic
center-of-mass motion for the atoms must be treated quantum mechanically;
for $L\ll L_{T}$, this motion can be treated classically. Typically, $L_{T}$
is of order of a few cm for a thermal beam. We restrict our discussion to
the classical scattering limit. For classical scattering, the {\em total}
atomic density is not modulated, as long as no state-selective mechanism
removes atoms from the system. In other words, the excited and ground state
population can be modulated following interaction with the fields, but the 
{\em sum} of these populations is not modified. If the excited state
population decays to the ground state following interaction with the field,
the grating structure disappears. There are several methods to maintain the
gratings following the atom-field interaction. The most direct method is to
use optical transitions, such as those in Ca, Mg, or Yb, with lifetimes
approaching a microsecond or longer. One can then probe the population {\em %
difference} grating following the atom-field interaction or selectively
ionize excited state atoms to leave a net atomic density grating that could
be deposited on a substrate.

For shorter-lived atomic transitions, there are several alternatives. One
possibility is to ionize some of the excited state atoms {\em during} the
atom-field interaction, which would leave a net atomic density grating.
Another possibility is to use atoms whose ground state consists of a
manifold of substates. When the excited state decays, the {\em total }ground
state density is not modulated, but that of specific substates may be
modulated. By selectively ionizing some of the ground state sublevels, one
again achieves a net atomic density grating. Still another alternative is to
use a beam of metastable atoms and drive transitions between the metastable
state and an excited state that can decay back to the ground state \cite
{metastable}. Following the atom-field interaction, there will be a grating
of metastable atoms that can be detected selectively.

In the case of shorter-lived excited states, one can run into problems
related to the effective pulse interaction time $T$ which corresponds to the
transit time of the atoms through the field interaction zone. Recall that
all frequencies are measured in units of $T^{-1}$. For appreciable gratings,
it is necessary that $\gamma T\lesssim 0.5$. If the fields are focused to a
spot size of 10 $\mu $ and the longitudinal atomic speed is $5\times 10^{4}$
cm/s, then $T\approx 20$ ns. For these parameters, one must use atoms whose
excited state lifetime exceeds 20 ns. It is also possible to limit the
atom-field interaction time by using pulsed fields. For transit times of
order 20 ns, the detunings and Rabi field strengths considered in this paper
correspond to frequencies in the MHz to GHz range. Such values do not pose
any serious experimental difficulties.

Assuming that we can produce a density grating, many methods are available
for imaging the gratings. Perhaps the most direct method is to detect the
grating immediately following the atom-field interaction. This scheme is
analogous to free-induction decay (FID) in coherent transient spectroscopy.
All harmonics in the beam are superimposed, and the grating begins to wash
out as a result of the transverse Doppler effect at distances $L_{b}=\lambda
/(2n\theta _{b})$, where $\theta _{b}$ is the atomic beam divergence. To
avoid overlap of the different harmonic components, echo techniques can be
used \cite{ampgrat}. With two interaction zones separated by a distance $%
L_{0},$ different harmonics are focused at different distances following the
second interaction zone. For example, one can optimize the $(2n)$th harmonic
at $L=2L_{0}$, the $(4n)$th harmonic at $L=3L_{0}/2,$ the $(6n)$th harmonic
at $L=4L_{0}/3$, etc. In this manner one is able to isolate and focus
harmonics higher than $2n$ which were created in the first field interaction
zone \cite{dubet1}. For example, one could create 4th harmonic in the first
interaction zone and 16th harmonic in the second interaction zone to
optimize the $12$th harmonic at $L=4L_{0}/3$.

Although FID and echo techniques can be used also for phase gratings \cite
{cahn}, we consider here only the focusing capabilities of phase gratings.
The phase grating can be thought to arise as the result of an ac Stark shift
potential that acts as a spatially modulated index of refraction for the
atoms and causes them to focus at a distance of order $L_{T}/A$, where $A$
is some effective pulse area in the problem associated with the creation of
the $\left( 2n\right) $th harmonic. In this manner, one can focus an atomic
beam to a series of lines (or dots if a 2-D geometry is utilized) having
spacing $\lambda /2n$.

In summary, we have outlined methods for generating and detecting spatially
modulated atomic density amplitude or phase gratings having period $\lambda
/2n$ using optical fields having wavelength $\lambda .$ All harmonics having
larger periods are suppressed by a proper choice of atom-field detunings.

\section{Acknowledgments}

This work is supported by the U. S. Office of Army Research under Grant No.
DAAG55-97-0113 and the National Science Foundation under Grant No.
PHY-9414020.

\end{multicols}
\end{document}